\documentstyle[12pt]{article}
\renewcommand
\baselinestretch{1.3}

\begin{document}

\def\beq{\begin{equation}}
\def\eeq{\end{equation}}
\def\bce{\begin{center}}
\def\ece{\end{center}}
\def\bea{\begin{eqnarray}}
\def\eea{\end{eqnarray}}
\def\ben{\begin{enumerate}}
\def\een{\end{enumerate}}
\def\ul{\underline}
\def\ni{\noindent}
\def\nn{\nonumber}
\def\bs{\bigskip}
\def\wt{\widetilde}
\def\rl{$\Real$}
\def\tu{\bigtriangleup}
\def\td{\bigtriangledown}
\def\brr{\begin{array}}
\def\err{\end{array}}
\def\y{\index}
\def\ri{{\rm i}}

%\hfill HUPD-93?

\hfill November 1993

\vspace*{3mm}

\begin{center}

{\LARGE \bf
Renormalization-group improved effective Lagrangian for interacting
theories in curved spacetime}

\vspace{4mm}

\renewcommand
\baselinestretch{0.8}
\medskip

{\sc E. Elizalde}
\footnote{E-mail: eli @ ebubecm1.bitnet} \\
Department E.C.M. and I.F.A.E., Faculty of Physics,
University of  Barcelona, \\ Diagonal 647, 08028 Barcelona, \\
and Blanes Center for Advanced Studies, C.S.I.C., 17300 Blanes,
Spain \\
 and \\
{\sc S.D. Odintsov}\footnote{E-mail: odintsov @ ebubecm1.bitnet} \\
Tomsk Pedagogical Institute, 634041 Tomsk, Russian Federation, \\
and Department E.C.M., Faculty of Physics,
University of  Barcelona, \\  Diagonal 647, 08028 Barcelona,
Spain \\

\renewcommand
\baselinestretch{1.4}

\vspace{5mm}

{\bf Abstract}

\end{center}

A method for finding the renormalization group (RG) improved
effective
Lagrangian
for a massive interacting field theory in curved spacetime is
presented.
As a particular example, the $\lambda \varphi^4$-theory is
considered and the RG improved effective Lagrangian is explicitly
found
up to second order in the curvature tensors. As a further
application,
the
curvature-induced phase transitions are discussed for both the
massive
and the massless versions of the theory.
The problems which appear when calculating the RG improved effective
Lagrangian for gauge theories are discussed, taking as  example
the asymptotically free SU(2) gauge model.

\vspace{4mm}

\newpage

\ni 1. \ The concept of effective potential turns out to be very
useful
in modern particle physics and quantum cosmology. The effective
potential formalism combined with the renormalization group (RG)
method is even more important in particle physics phenomenology.
Specifically, starting from the one-loop effective potential
in renormalizable theories one can obtain the RG improved effective
potential \cite{1,2}, by summing all the leading (or subleading)
logarithms. Thus, starting from a limited range of the background
scalars, one can actually extend this range to other values, what
renders this formalism very useful in practice \cite{1,2}. Using
the RG effective potential in a modified form \cite{3,4},
estimations on the absolute stability of the electroweak vacuum in
the standard model have been recently discussed \cite{3}.
{}From a different side, the effective potential in quantum field
theory in curved spacetime has been also widely discussed recently
(for a general review, see \cite{5}), starting from Ref. \cite{6},
where the possibility of curvature-induced symmetry breaking (or
restoration) was realized. The explicit calculations of the
one-loop effective potentials for some simple theories, such as the
$\lambda \varphi^4$-theory or scalar electrodynamics have been done
on a number of different background spaces (for a list of
references, see \cite{5}). On an arbitrary, non-static curved
spacetime we need a generalization of the concept of effective
potential,
i.e., the effective Lagrangian which can be calculated as
an expansion of the effective action up to terms of some
 fixed-order  on the curvature tensors. The effective Lagrangian up
to quadratic terms in the curvature tensors has been found both for
the $\lambda \varphi^4$-theory \cite{8} and for scalar
electrodynamics (see \cite{5}, Chap. 5).

Recently \cite{7} the RG improved effective potential in the linear
curvature approximation, for an arbitrary, renormalizable massless
gauge theory (including GUTs) in curved spacetime has been
presented, thus generalizing Coleman-Weinberg's approach \cite{1,2}
corresponding to flat space. The curvature-induced phase
transitions, based on the form of this potential, have been
discussed in detail and the possibility of such a phase transition
in the SU(5) GUT has been shown. This effective potential may be
relevant for early universe considerations, in particular, for the
discussion of the inflationary universe (for a review and an
introduction, see \cite{10,12}). There, most of the actual
studies have been carried out for a flat-space potential.

The next problem which shows up naturally is to generalize the
approach of Ref.  \cite{7} to the case of massive interacting
theories in general curved spacetime. (Note that in such a
situation we need the effective Lagrangian up to, at least, second
order terms in the curvature tensors). In this work we introduce
such
a generalization, namely we present an explicit method in order to
obtain the RG improved effective Lagrangian for massive interacting
theories. Notice that this is quite remarkable, since the RG
improved effective Lagrangian gives the result beyond one loop,
i.e., it manages to sum all the logarithms of perturbation theory. Note
that already two-loop
calculations of the  $\beta$-functions in curved space
are very hard to do \cite{9}. As an example, the case of the
$\lambda \varphi^4$-theory is studied in detail, where we find the RG
improved effective Lagrangian up to second order in the curvature
tensors.
 \bs

\ni 2. \ Let us consider the self-interacting scalar field theory
with the Lagrangian
\beq
L_m = \frac{1}{2} g^{\mu\nu} \partial_\mu \varphi  \partial_\nu
\varphi - \frac{1}{2} (m^2- \xi R) \varphi^2 - \frac{\lambda}{24}
\varphi^4.
\label{1}
\eeq
In order to have a renormalizable theory in curved spacetime, as
usually, we must add to (\ref{1}) the Lagrangian corresponding to
the external gravitational field
\beq
L_G = \Lambda + \kappa R + a_1 R^2 + a_2 C_{\mu\nu\alpha\beta}^2 +
a_3 G +a_4 \Box R.
\eeq
then, the whole multiplicatively-renormalizable theory  is described by
\beq
L=L_m+L_G \label{3}
\eeq
(for details and an introduction, see \cite{5}).

We will be interested in the study of the effective action for the
theory (\ref{3}). Since it turns out to be impossible to calculate
the effective action on curved space-time even in the one-loop
approximation, we will study the one-loop effective Lagrangian of
the theory, i.e., the expansion of the one-loop effective action
(up to quadratic terms in the curvature tensors). Such effective
Lagrangian, using some specific renormalization conditions, has been
already found in Ref. \cite{8}.
It should be noted that this effective Lagrangian (to the order we are
interested in), in addition to the terms explicitly written in
(\ref{3}) should also contain the term
\beq
\Delta L = a_5 \Box \Phi^2,
\eeq
which is a trivial total derivative at the tree level, but which
becomes essential at the quantum level.

It is well known that a theory with the action (\ref{3}) is
multiplicatively renormalizable in curved spacetime. Then, the
effective action (and also the effective Lagrangian) satisfies the
standard RG equation
\beq
\left( \mu \frac{\partial}{\partial \mu} + \beta_i
\frac{\partial}{\partial \lambda_i} - \gamma \Phi
\frac{\partial}{\partial \Phi} \right) L_{eff} (\mu,\lambda_i,
\Phi)=0,
\eeq
where  $\lambda_i =(\xi, \lambda, m^2, \Lambda, \kappa, a_1,
\ldots, a_5) $ are all the coupling constants, including also
dimensional coupling constants, $\beta_i$ are the corresponding
$\beta$-functions and $\gamma$ is the $\gamma$-function of the
scalar field. Notice that from the point of view of the background
field method $\Phi$ is the background scalar field.

A solution of the RG equation can be easily found by the method of
the characteristics:
\beq
L_{eff}  (\mu,\lambda_i,\Phi)= L_{eff}  (\mu (t),\lambda_i (t),\Phi
(t)),
\eeq
where $\Phi (t)= \Phi \exp \left( - \int_0^t \gamma (t') \, dt'
\right)$, $\mu (t) =\mu \, e^t$, and $\lambda_i (t)$ are the
effective coupling constants defined as the solutions of the
equation
\beq
\frac{d \lambda_i (t)}{dt} = \beta_i (\lambda_i (t)), \ \ \ \ \
\lambda_i (0) = \lambda_i.
\eeq
Now, we need the one-loop $\beta$-functions of the theory. (These
$\beta$-functions have been calculated up to two-loop order in flat
space \cite{3} and also in curved space \cite{9}).

Explicitly, the  one-loop $\beta$-functions are
\bea
&& \beta_\lambda = \frac{3\lambda^2}{(4\pi)^2}, \ \ \   \beta_{m^2}
= \frac{\lambda m^2}{(4\pi)^2}, \ \ \  \gamma =0, \ \ \  \beta_\xi
= \frac{\lambda (\xi - 1/6)}{(4\pi)^2}, \nn \\
&& \beta_\Lambda = \frac{m^4}{2(4\pi)^2}, \ \ \   \beta_{\kappa} =
\frac{m^2 (\xi -1/6)}{(4\pi)^2},  \ \ \  \beta_{a_1} = \frac{(\xi
-
 1/6)^2}{2(4\pi)^2},   \ \ \  \beta_{a_2} = \frac{1}{120(4\pi)^2},
\nn \\
&&  \beta_{a_3} =- \frac{1}{360(4\pi)^2},  \ \ \  \beta_{a_4} = -
\frac{\xi - 1/5}{6(4\pi)^2},   \ \ \  \beta_{a_5} = -
\frac{\lambda}{12(4\pi)^2}.
\label{8}
\eea
Starting from the tree-level effective Lagrangian, we will find the
RG improved effective Lagrangian in the following form
\bea
L_{eff} &=&  \frac{1}{2} g^{\mu\nu} \partial_\mu \Phi  \partial_\nu
\Phi - \frac{1}{2} [m^2(t)- \xi (t) R] \Phi^2 - \frac{\lambda
(t)}{24}\Phi^4 + a_5(t) \Box \Phi^2 + \Lambda (t) \nn \\
&&+ \kappa (t) R + a_1 (t) R^2 + a_2 (t) C^2_{\mu\nu\alpha\beta}
+a_3(t) G + a_4(t) \Box R.
\label{9}
\eea
The effective coupling constants are defined by Eqs. (\ref{8}) to
be
\bea
&&\lambda (t) = \lambda \left( 1- \frac{3\lambda t}{(4\pi)^2}
\right)^{-1}, \ \ \ m^2(t) = m^2  \left( 1- \frac{3\lambda
t}{(4\pi)^2} \right)^{-1/3}, \nn \\
&&\xi (t) = \frac{1}{6}+ \left( \xi - \frac{1}{6} \right)  \left(
1- \frac{3\lambda t}{(4\pi)^2} \right)^{-1/3}, \ \ \ \Lambda (t) =
\Lambda - \frac{m^4}{2\lambda} \left( 1- \frac{3\lambda
t}{(4\pi)^2} \right)^{1/3} + \frac{m^4}{2\lambda}, \nn \\
&&\kappa (t) = \kappa -\frac{m^2}{\lambda} \left( \xi - \frac{1}{6}
\right) \left[ \left( 1- \frac{3\lambda t}{(4\pi)^2} \right)^{1/3}
-1
\right],
\nn \\ && a_1 (t) = a_1 -\frac{1}{2\lambda} \left( \xi -
\frac{1}{6}
\right)^2 \left[ \left( 1- \frac{3\lambda t}{(4\pi)^2}
\right)^{1/3}
-1\right],
\nn \\ && a_2 (t) = a_2 +\frac{t}{120 (4\pi)^2}, \ \ \  a_3 (t) =
a_3
-
\frac{t}{360(4\pi)^2}, \nn \\ && a_4 (t) = a_4 +\frac{t}{180
(4\pi)^2} +\frac{1}{3\lambda} \left( \xi - \frac{1}{6} \right)
\left[ \left( 1- \frac{3\lambda t}{(4\pi)^2} \right)^{2/3}
-1\right],
\nn \\ && a_5 (t) = a_5 +\frac{1}{36} \ln  \left( 1- \frac{3\lambda
t}{(4\pi)^2} \right).
\eea
A few remarks are in order. A convenient choice of the parameter
$t$ (compare with flat space \cite{3,4}) is
\beq
\mu^2 (t) = \mu^2 e^{2t} = m^2-  \left( \xi - \frac{1}{6} \right)
R + \frac{\lambda}{2} \Phi^2.
\label{11}
\eeq
Such a combination appears naturally in a direct calculation of the
one-loop efective Lagrangian. Then, the RG improved effective
Lagrangian (\ref{9}) (summing all the logarithms) in the limit
$\lambda <<1$,  $|\lambda t| <<1$, easily reproduces the results of
the much more tedious one-loop calculation \cite{8} if the same
renormalization conditions are imposed. However, notice that
(contrary to what happens in the non-improved case) Eq. (\ref{9})
is actually valid for all $t$ for which $L_{eff}$ does not diverge
(this is the improvement).

If we consider the flat space limit of (\ref{9}), i.e. $R=0$ and
also take $\Phi=$ const., then Eq. (\ref{9}) reduces to the RG
improved effective potential
\beq
- L_{eff} = V = \frac{\lambda (t)}{24} \Phi^4 + \frac{m^2(t)}{2}
\Phi^2 - \Lambda +\frac{m^4}{2\lambda} \left( 1- \frac{3\lambda
t}{(4\pi)^2} \right)^{1/3}- \frac{m^4}{2\lambda},
\eeq
with $t=(1/2) \ln [(m^2+ \lambda \Phi^2/2)/\mu^2]$.

This is the standard RG improved effective potential with the
effective cosmological constant $\Lambda (t)$ precisely of the form
introduced in Ref. \cite{4} (see also the discussion in \cite{3}).
Thus we have the natural curved-space interpretation as the
effective cosmological constant for the RG improved potential at
zero background field.
\bs

\ni 3. \ Let us consider now some application of the effective
Lagrangian
(\ref{9}). Choosing for simplicity the constant curvature space
$R_{\mu\nu} = (R/4) g_{\mu\nu}$ (then, naturally, $\Phi =$ const.),
and working in the linear curvature approximation, we get
\beq
 L_{eff} = -\frac{\lambda (t)}{24} \Phi^4 - \frac{1}{2}
\left[m^2(t)-\xi (t)R \right] \Phi^2 + \Lambda (t) + \kappa (t) R.
\label{13}
\eeq
Already the classical Lagrangian (\ref{1}) exhibits the possibility
of spontaneous symmetry breaking \cite{6}
\beq
\Phi^2 = \frac{2}{\lambda} (\xi R-m^2 ),
\eeq
for $\xi R >m^2$.

We now investigate the phase structure of the RG improved potential
(\ref{13}). It is nowadays common to think that the very early
universe experienced several phase transitions before it reached
its present state. Some models of the inflationary universe are
based on the temperature phase transition \cite{10,12}. It
is also possible that a phase transition could be induced by the
strong gravitational field existing in the very early universe.
We will discuss only the possibility of a first order phase
transition, where the order parameter, $\Phi$, experiences a
quick change for some critical value, $R_c$, of the curvature. For
simplicity, we will put $\Lambda =\kappa =0$ in (\ref{13}), since
the general situation is just a rescaling of the constant part of
the potential.

One can rewrite (\ref{13}) as follows
\beq
-\frac{ L_{eff}}{\mu^4} = \frac{V}{\mu^4} = \frac{\lambda (t)
x^2}{24} + \frac{1}{2} \left[m^2(t)-\xi (t)y \right] x +
 \left[ \left( 1- \frac{3\lambda
t}{(4\pi)^2} \right)^{1/3} -1\right] \left[
\frac{\wt{m}^4}{2\lambda}+
\frac{\wt{m}^2(\xi -1/6)y}{\lambda} \right],
\label{15}
\eeq
where \[ x = \frac{\Phi^2}{\mu^2}, \ \ \ \wt{m} = \frac{m}{\mu},
\ \ \ y = \frac{R}{\mu^2}, \ \ \ t = \frac{1}{2} \ln \left[
\frac{\lambda}{2}x +\wt{m}^2 - \left( \xi - \frac{1}{6} \right) y
\right]. \]
\bs

\ni 4. \ We can now extend the analysis of critical points, that
was performed in Ref. \cite{7} for the massless case, to the
general action (\ref{15}) corresponding to the massive case. Let us
recall that the critical
 parameters, $x_c$, $y_c$, corresponding to the first-order
phase transition are found from the conditions
\beq
V(x_c,y_c)=0, \ \ \ \ \left. \frac{\partial V}{\partial x}
\right|_{x_c,y_c} =0, \ \ \ \ \left. \frac{\partial^2 V}{\partial
x^2} \right|_{x_c,y_c} >0.
\label{cp}
\eeq
For the RG improved potential
they lead to some transcendental equations which cannot be solved
analytically.
We shall be concerned with first-order phase transitions where the
order parameter $\Phi$ experiences a quick change for some
critical value, $R_c$, of the curvature. Let us again recall first
the case of the $\lambda \Phi^4$ theory without mass, where
one has \cite{7}
\beq
\frac{V}{\mu^4} = \frac{\lambda x^2}{4! \left( 1- \frac{3\lambda
\ln x}{32\pi^2} \right)}- \frac{1}{2} \epsilon y x
\left[ \frac{1}{6} + \left( \xi - \frac{1}{6} \right) \left( 1-
\frac{3\lambda \ln x}{32\pi^2} \right)^{-1/3} \right],
\label{m-o}
\eeq
with $x=\Phi^2/\mu^2$,  $y=|R|/\mu^2$, and $\epsilon =$ sgn $R$.
The analysis of extrema of (\ref{m-o}) yields the following result.
One looks for critical points
$(x_c, y_c)$ defined by the simultaneous conditions (\ref{cp}).
The first two equations (\ref{cp}) yield, for the massless
potential (\ref{m-o}),
\beq
y_c=\frac{\epsilon \lambda x_c}{ 2 u \left[ 1+(6\xi-1)
u^{-1/3}\right]}, \ \
 \frac{32\pi^2}{3\lambda}\, u -\frac{1}{3\left[ 1+ (6\xi
-1)u^{1/3}\right]} +1=0,
\ \ u\equiv 1- \frac{3\lambda \ln x_c}{32\pi^2}.
\label{ce18}
\eeq
 The following models are particularly interesting.

\ni {\it (a)  Chaotic inflationary model.}
For  $\lambda =
10^{- 13}$ and $\xi =0$, both for positive $\epsilon =1$ and negative
$\epsilon =-1$ curvature,
a critical value appears, which
lies close to the pole ($x_p$):
\beq
x_c=\exp \left(- \frac{2}{3} \, 10^{-15} \right) \, x_p, \ \ \ \
y_c=- \epsilon \, 10^{-3} \, x_c, \ \ \ \ x_p = \exp \left(
\frac{32\pi^2}{3\lambda} \right).
 \eeq
This point is a minimum of (\ref{m-o}),
that is, all three equations (\ref{cp}) are indeed
satisfied.
 (Quite on the contrary, the
one-loop effective action  does
not yield any phase transition, the solutions  being
$x_c=0$, $y_c=0$, and
$x_c=1$, $y_c= \epsilon \, 10^{17} \, x_c$).

\ni {\it (b)  Variable Planck-mass model.}
 For  $\lambda$ we take a typical
value corresponding to particle physics models, e.g. $\lambda = 0.05$.
For
$\xi$ we choose two different values: (b1) $\xi = - 10^4$
(which actually corresponds to Ref. \cite{sbb})
 and (b2)  $\xi
= 1/6$, respectively.

 In case (b1), the critical point corresponding to
(\ref{m-o}) is obtained for
\beq
x_c= e^{2/3} \, x_p, \ \ \ y_c= -5 \cdot 10^{-5} \, x_c,
\eeq
both for positive and for negative curvature. (For the one-loop action,
the only solution is again the trivial one $x_c=y_c=0$).

 In  case (b2), the critical point for the RG improved action (massless
case) is at
\beq
x_c= e \, x_p, \ \ \ y_c= -\epsilon \, 50 \, x_c,
\eeq
which is {\it not} consistent with our approximation $x_c>>y_c$.
(For the one-loop effective action
 $x_c=y_c=0$ is again the only solution).

Guided by the results of the analysis above corresponding to the
massless case ---and in order to simplify the more involved one of the
massive case---
it is natural to start by rescaling the variables and the potential in
(\ref{15}) as follows:
\bea
&& \bar{x} = x \, \exp \left( -\frac{32 \pi^2}{3\lambda} \right),
\ \ \  \bar{y} = y \, \exp \left( -\frac{32 \pi^2}{3\lambda}
\right),  \ \ \ \bar{m}^2 = \wt{m}^2 \, \exp \left( -\frac{32
\pi^2}{3\lambda} \right),  \nn \\
&& \bar{u} = - \frac{3\lambda}{32 \pi^2} \bar{t},  \ \ \ \bar{t} =
\ln  \left[ \frac{\lambda}{2}\bar{x} +\bar{m}^2 - \left( \xi -
\frac{1}{6} \right) \bar{y} \right], \ \ \  \bar{V} =
\frac{V}{\mu^4} \, \exp \left( -\frac{64 \pi^2}{3\lambda} \right),
\label{resc}
\eea
so that we obtain the simplified expression
\beq
\bar{V} = \frac{\lambda \bar{x}^2}{24 \bar{u}} + \frac{\bar{x}}{2}
\left\{ \bar{m}^2 \bar{u}^{-1 /3} - \bar{y}  \left[
\frac{1}{6} + \left( \xi - \frac{1}{6} \right)  \bar{u}^{-1 /3}
\right] \right\} + \frac{\bar{m}^2}{\lambda} \left( \bar{u}^{1/3} -1
\right) \left[  \frac{\bar{m}^2}{2} +  \left( \xi - \frac{1}{6}
\right) \bar{y} \right].
\label{mas1}
\eeq

How does the introduction of mass change the results given above
corresponding to the massless case? A
careful analysis of (\ref{mas1}) leads to the following conclusions. The
changes in the equations for the critical points can be absorbed by
a shift in the $y$-variable proportional to the mass, namely
\beq
\bar{y}  \longrightarrow \bar{y} - \frac{\bar{m}^2}{\xi - 1/6}.
\eeq
This is also valid for the second case (b1) above, but certainly not
 for the third case (b2).
 Let us recall, though, that this is the only situation in which we do
not obtain a critical point consistent with the approximation in
which we are working, already for the massless model. This fact is made
no better through
the introduction of mass. Again, concerning the first two cases, and
for reasonable values of the mass (i.e., $\bar{m}^2 < 1$) we do not get
a substantial change of the situation described for the massless
case. However, the nontrivial critical points do not show up any
more for higher values of the mass  ($\bar{m}^2 > 1$, recall the
rescaling (\ref{resc})). Thus we see that for positive $\xi > 1/6$ a
non-zero (positive) mass leads to a smaller critical curvature. Also,
for negative $\bar{m}^2$ what one gets is an increase of the critical
curvature, which gets very large near $\xi = 1/6$.

Notice, moreover, that in order to include higher-order terms in the
above
study one has to do already a very tedious numerical analysis for
different values of the parameters.

The higher-order terms in the
effective Lagrangian may be important in some different contexts.
For example, starting from the RG  improved effective
potential, taking into account second-order curvature terms, we may
expand it on $\Phi$. In this case the typical form of the potential
will be
\beq
V= \Lambda_{eff} + m_{eff}^2 \Phi^2 + {\cal O} (\Phi^4),
\eeq
where $m_{eff}^2$ includes all the curvature terms up to second
order. Then, it is easy to estimate the influence of the
second-order curvature terms
 on the symmetry breaking or restoration of
the theory, for different types of constant curvature spaces, and
to compare it with the corresponding results for the non-improved
effective potential \cite{8}.

Another interesting question is connected with the back reaction
problem. One can use the RG improved effective Lagrangian to study
Einstein equations with quantum corrections of interacting matter
fields. However, this analysis involves (again) tedious numerical
calculations, owing to the high non-linearity of the problem.
 \bs

\ni 5. \
Finally, let us discuss the generalization of the above method to
multiple-mass cases. As an explicit example we shall consider the
massive gauge theory based on the gauge group SU(2) with one
multiplet of scalars ($\varphi^a$, $a=1,2,3$), taken in the adjoint
representation of SU(2), and one or two multiplets of spinors taken
in the adjoint representation too (for details of this model in an
asymptotically free regime, see \cite{13}). This theory is
asymptotically free for all  coupling constants, and for the case
of one spinor multiplet it is also asymptotically conformally
invariant.

In principle, one can apply the above procedure in order to get the
RG improved effective potential of the theory. However, now some of
the effective masses are different:
\beq
m_B^2 =m^2 - \left( \xi - \frac{1}{6} \right) R + c_1\lambda
\Phi^2 + c_2 g^2 \Phi^2, \ \ \ \ \ M_F^2 = h^2 \Phi^2 +c_3R, \ \ \ \ \
M_{\mu\nu}^2 \simeq \delta_{\mu\nu} (g^2 \Phi^2 + c_4 \lambda
\Phi^2) + c_5R_{\mu\nu}, \eeq
where $c_1, \ldots, c_5$ are some constants, and $\Phi^2 =
\varphi^a
\varphi^a$. It turns out that it is impossible to choose $t$ in
(\ref{11}) for the theory under discussion in the same unique way
as for the $\lambda \varphi^4$-theory above. Of course, choosing
$t=\ln (\Phi /\mu)$ we always will have the correct behavior of
$L_{eff}$ at large $\Phi$ but not at all scales. The question is,
however, if we are able to construct the more exact, RG improved
effective Lagrangian, not just its asymptotic form. The answer is
certainly positive.

Indeed, working in the asymptotically free regime
\[  \lambda (t) = k_1 g^2(t), \ \ \ h^2(t) =k_2 g^2(t), \ \ \
g^2(t) = g^2 \left[ 1+ \frac{a^2 g^2 t}{(4\pi)^2} \right]^{-1}, \]
where $k_1, k_2$ and $a^2$ are given in \cite{13}, one may
construct the RG improved $L_{eff}$ as above, and then choose $t=
\frac{1}{2} \ln (h^2\Phi^2 /\mu^2)$. Then, one has (for simplicity,
we write the RG improved effective potential in the linear
curvature approximation only)
\bea
V_{eff} &=& \frac{1}{4!} \Phi^4 f^4(t) k_1 g^2(t) - \frac{1}{2} R
\Phi^2 f^2(t) \left[ \frac{1}{6} +  \left( \xi - \frac{1}{6}
\right)  \left( 1+ \frac{a^2 g^2 t}{(4\pi)^2}
\right)^{(5k_1/3+8k_2-12)/a^2} \right] \nn \\
&& -  \frac{1}{2}  \Phi^2 f^2(t) m^2 \left( 1+ \frac{a^2 g^2
t}{(4\pi)^2} \right)^{(5k_1/3+8k_2-12)/a^2}- \Lambda (t) - \kappa (t) R,
\eea
where
\[ f(t) =  \left( 1+ \frac{a^2 g^2 t}{(4\pi)^2} \right)^{(6-
4k_2)/a^2}, \ \ \
\Lambda (t) = \Lambda + \frac{3m^4}{2 a^2g^2 A} \left[ \left( 1+
\frac{a^2 g^2 t}{(4\pi)^2} \right)^A -1 \right],  \]
\[
\kappa (t) = \kappa + \frac{3m^2 (\xi - 1/6)}{a^2g^2 A} \left[ \left(
1+ \frac{a^2 g^2 t}{(4\pi)^2} \right)^A -1 \right], \ \ \
A = \left( \frac{5}{3} \kappa_1 + 8 \kappa_2 - 12 \right) \frac{1}{a^2}
+1.   \]
This potential is very similar (forgetting about the mass, cosmological
and Newton terms) to the one of Ref. \cite{7} corresponding to the
massless
case. However, this means that we get $L_{eff}$ (or $V_{eff}$) only in
the region $h^2\Phi^2 >m^2$. In order to obtain the RG improved
effective Lagrangian we have to find it in other regions.
 This is not so
straightforward and demands some careful considerations (using, for
instance, the multiscale RG of Einhorn and Jones \cite{2}), that will be
presented in future work.

\vspace{5mm}

\noindent{\large \bf Acknowledgments}

SDO would like to thank the members of the Dept. ECM, Barcelona
University, for the kind hospitality.
This work has been  supported by DGICYT (Spain) and by CIRIT
(Generalitat de Catalunya).


\newpage

\begin{thebibliography}{99}

\bibitem{1} S. Coleman  and E. Weinberg,  Phys. Rev. {\bf D7}
(1973) 1888.

\bibitem{2} M.B. Einhorn and D.R.T. Jones, Nucl. Phys. {\bf
B211} (1983) 29;  {\bf B230} (1984) 261; G.B. West,
Phys.Rev. {\bf D27} (1983) 1402; K. Yamagishi, Nucl. Phys.
{\bf B216} (1983) 508; M. Sher, Phys. Rep. {\bf 179} (1989)
274.

\bibitem{3}  C. Ford, D.R.T. Jones, P.W. Stephenson and
M.B. Einhorn, Nucl. Phys. {\bf B395} (1993) 17.

\bibitem{4} B. Kastening, Phys. Lett. {\bf B283} (1992) 287.

\bibitem{5} I.L. Buchbinder, S.D. Odintsov and I.L. Shapiro,
{\sl
Effective Action in Quantum Gravity}, IOP Publishing, Bristol
and
Philadelphia, 1992.

\bibitem{6} G.M. Shore, Ann. Phys. NY {\bf 128} (1980) 376.

\bibitem{7} E. Elizalde  and S.D. Odintsov, Phys. Lett. {\bf B303}
(1993) 240.

\bibitem{8} K. Kirsten, G. Cognola and L. Vanzo, Phys. Rev. {\bf D48}
(1993) 2813.

\bibitem{9} I. Jack and H. Osborn, Nucl. Phys. {\bf B249} (1985)
472; ibid. {\bf B253} (1985) 323.

\bibitem{10} E.W. Kolb and M.S. Turner, {\it The Early
Universe},
Addison-Wesley, 1990.

\bibitem{12} A.D. Linde, {\it Particle Physics and
Inflationary
Cosmology}, Contemporary Concepts in Physics, Harwood
Academic, New
York, 1990.

\bibitem{13} B.L. Voronov and I.V. Tyutin, Yad. Fiz. [Sov. J. Nucl.
Phys.] {\bf 23} (1976) 664; I.L. Buchbinder and S.D. Odintsov,
Yad. Fiz. [Sov. J. Nucl. Phys.] {\bf 40} (1984) 1338.

\bibitem{sbb} D. Salopek, J.R. Bond and J.M. Bardeen, Phys. Rev. {\bf
D40} (1989) 1753.

\end{thebibliography}
\end{document}